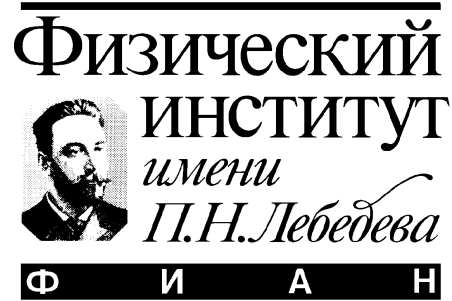

A.V. BAGULYA, O.D. DALKAROV, M.A. NEGODAEV, A.S. RUSETSKII, A.P. CHUBENKO, V.G. RALCHENKO, A.P. BOLSHAKOV

# CHANNELING EFFECT IN POLYCRYSTALLINE DEUTERIUM-SATURATED CVD DIAMOND TARGET BOMBARDED BY DEUTERIUM ION BEAM

# CHANNELING EFFECT IN POLYCRYSTALLINE DEUTERIUM-SATURATED CVD DIAMOND TARGET BOMBARDED BY DEUTERIUM ION BEAM

Bagulya A.V., Dalkarov O.D., Negodaev[*] M.A., Rusetskii[*] A.S., Chubenko A.P.

*Lebedev Physical Institute RAS, Moscow, Russia*

Ralchenko V.G., Bolshakov A.P.

*General Physical Institute RAS, Moscow, Russia*

## ABSTRACT

At the ion accelerator HELIS at the LPI, the neutron yield is investigated in DD reactions within a polycrystalline deuterium-saturated CVD diamond, during an irradiation of its surface by a deuterium ion beam with the energy less than 30 keV. The measurements of the neutron flux in the beam direction are performed in dependence on the target angle, $\beta$, with respect to the beam axis. These measurements are performed using a multichannel detector based on $^3$He counters. A significant anisotropy in neutron yield is observed, it was higher by a factor of 3 at $\beta=0°$ compared to that at $\beta = \pm 45°$. The possible reasons for the anisotropy, including ion channeling, are discussed.

[*] Corresponding author. E-mail: negodaev@lebedev.ru
[*] Corresponding author. E-mail: rusets@lebedev.ru

The HELIS facility at the LPI operates with continuous ion beams with currents up to 40 mA and energies up to 50 keV. This multi-purpose accelerator addresses the wide spectrum of physics experiments, like e.g. light nuclei collisions at energies of several keV, investigation of elementary and collective processes in ion-beam plasma and studies of the beam-target interactions using different materials with modification of the properties of the latter through ion-beam spattering of the thin-film coatings. Nowadays, at HELIS, the interactions of the deuterium beam with deuterium-enriched fixed targets [1-4].

In this paper, the recent results obtained at the HELIS facility are presented. The neutron yield in the DD-reaction at the deuterium-enriched CVD diamond is measured as a function of the beam incident angle. The neutrons originating in the reaction

$$d+d \rightarrow n(2.45 \text{ MeV}) + {}^3\text{He} \ (0.8 \text{ MeV}) \qquad (1)$$

are identified using the multichannel neutron detector with filled with $^3$He. The experimental setup is schematically represented in Fig.1

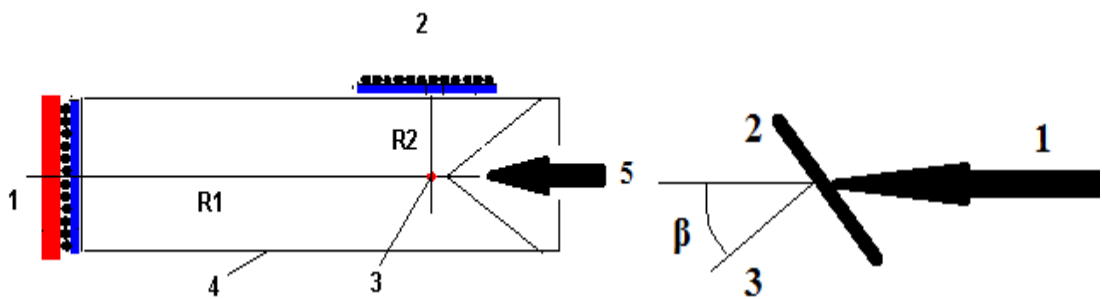

Fig 1. Left panel: the $^3$He detector setup at HELIS, representing the first (1) and the second (2) $^3$He-counter groups with radii R=85 cm and R=38 cm, respectively. The target is placed at (3) inside the HELIS beam pipe (4). The ion beam direction is indicated by (5). Right panel: The beam scattering angle beta: the beam is indicated by (1), the target is shown by (2), the normal to the target surface (3).

The first $^3$He-detector group consists of counters with radiators of 5 cm paraffin and of 3 cm organic glass. It is placed at R1=85 cm away from the target and identifies neutrons coming along the initial beam direction. The second group of $^3$He counters with 3cm organic glass radiators is placed at R2=38 cm and detects neutrons emitted transverse to the beam direction. The calibration of the $^3$He detectors is performed, using instead of the target the neutron source Cf-252, with the 4 π-radiation activity of $3 \times 10^4$ n/second. The fast-neutron registration efficiency of the both detector groups was determined to 0.13 % along and transverse to the beam direction, respectively.

In our previous investigations [1,2] of DD reaction in the crystal targets (Pd, Ti), an anisotropy was observed: the neutron flux along the beam direction was higher than that in the transverse direction. This effect could be explained by a DD-reaction proceeding not with creation of a compound nucleus $^4$He, but through partial penetration of the deuteron in the nucleus. Particularly large anisotropy was observed using a polycrystalline CVD diamond target.

The CVD-diamond samples were produced in the plasma-chemical reactor STS-100 using microwave discharge (frequency of 2.45 GHz microwave power up to 5 kW) in gas mixtures $CH_4$-$N_2$-$O_2$ and have a thickness of 400 microns. As a substrate, a silicon plate with a thickness of 3 mm and of a diameter of 57 mm, was used. The parameters of the growing process are as follows: the microwave power of 3,3 kW; the chamber pressure of 90 Torr; the substrate temperature of 920 $^o$C; the total gas flow of 0.5 l/min; the concentration of methane and oxygen 10% and 0.9%, respectively. Under these conditions, black diamond films were obtained, with numerous structural defects in the crystallites, such as twins and amorphous area with a size of about 1 nm. In the literature this material is referred to as a 'black diamond'. As the next step, the diamond film was separated from the substrate by etching of silicon in a mixture of hydrofluoric nitric and acetic acids, and cut by the Nd:YAG laser (pulse duration of about 10 nm and the frequency of 10 kHz) into discs with a diameter of 18 mm each. The graphite formed at the edges of the diamond disk during the cutting process, were removed by an oxidation in air at a temperature of 580 $^o$C during 1 hour. The structure of polycrystalline diamond is anisotropic and not homogeneous (Fig. 2a). The crystallites are growing in the form of columns, perpendicular towards the surface. With increasing the film thickness, the column diameter increases. The size of crystallites d increases from the 1 micron in the layer close to the substrate to about 100 micron on the growth side. The growth surface of the sample shows a clear crystallite structure, as shown in Fig. 2b. The quadratic fragments have typically the lateral size of d≈50 microns and are oriented in the plane of the film.

Hydrogen is the main admixture in the black CVD diamond, as observed in the excitations of C-H in the IR absorption spectra in the range of 2700-2900 cm$^{-1}$. The contribution of the C-H connections might reach to the order of 1000 ppm. Bind hydrogen is located on the grain boundaries and in the defects, decorating them. Nitrogen contribution is measured to be of the order of 1-10 ppm. The silicon admixture of about 10 ppm is mainly contained in the layer adjacent to the substrate. The CVD-diamond was saturated with deuterium through electrolysis in a 0.3M solution of LiOD in D$_2$O, using the diamond samples as a cathode together with a Pt anode. The voltage of 50 V was applied with the current density of 20-30 mA/cm$^2$. A penetration of the CVD diamond by about $10^{20}$ deuterium atoms could be concluded from the measurements of the electrolysis current and of the sample mass increase.

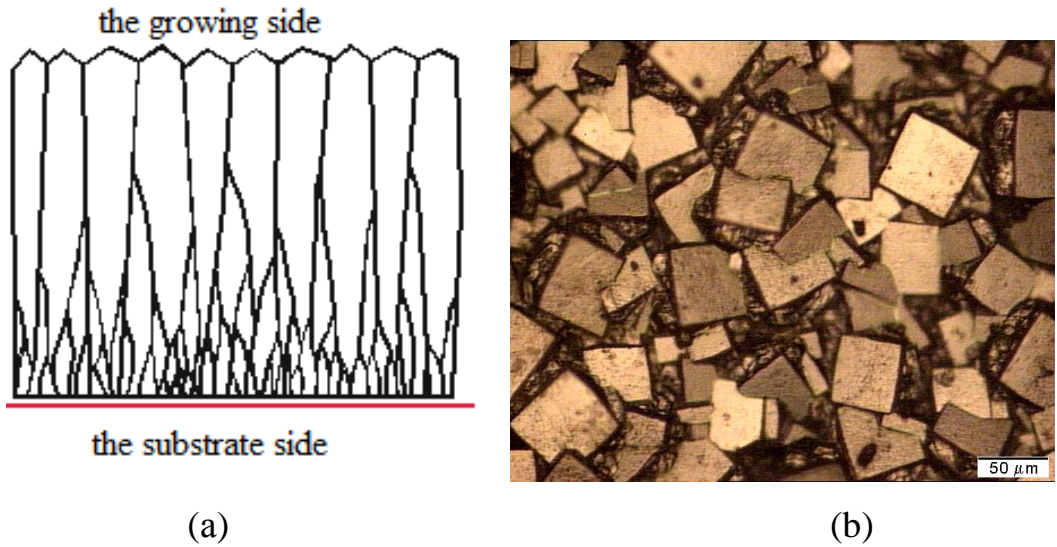

(a)                                   (b)

Fig. 2. The schematic view of the structure of columnar polycrystalline diamond film (a), and the structure of the growth surface, as obtained with an optical microscope (b).

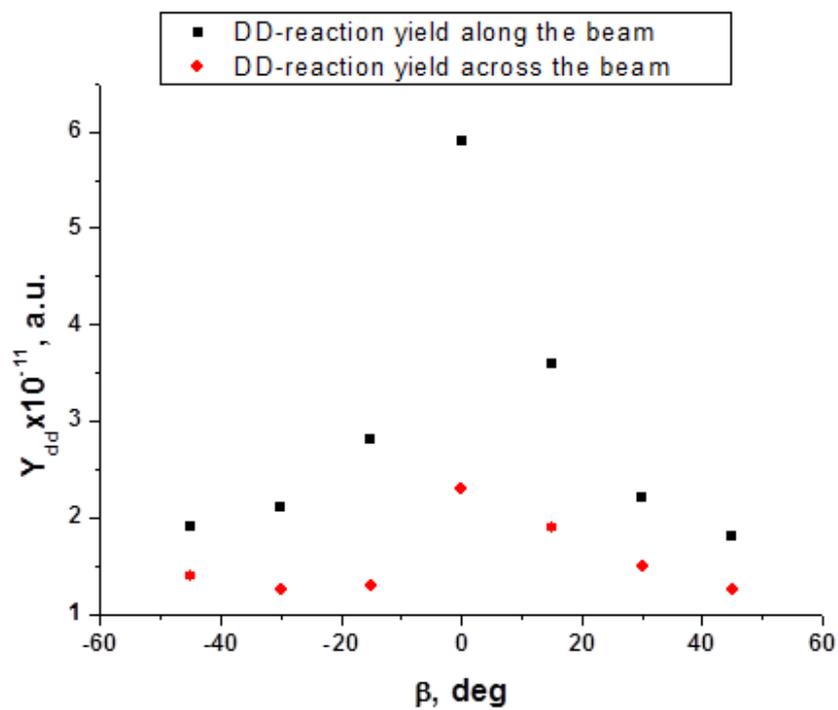

Fig. 3. The neutron yield obtained with the CVD-diamond sample as a function of the angle between the beam and the target plane norm, measured longitudinally (black squares) and transverse (red diamonds) directions with respect to the ion beam.

The target was installed in the water-cooled target-holder, supplied by a rotator varying the target angle β without any distortion of the vacuum conditions. At the HELIS facility, the target was irradiated with the deuterium ion beam with the energy of $E_d$=20 keV and the current of 50-60 μA.

The outcoming neutron flux, produced in the DD reaction, was measured in the longitudinal and transverse direction with respect to the beam axis by using a multichannel neutron detector based on $^3$He counters. The relative yield of the DD reaction was determined as $Y_{dd} = n_n/(S \times I_d)$, where $n_n$ is the longitudinal or transverse neutron flux, S is the irradiated area of the target and $I_d$ denotes the ion beam current.

The observed neutron yield measured as a function of the angle β between the beam direction and the norm to the target plane is shown in Fig. 3. It is observed, that the crystalline structure and the orientation of the sample with respect to the beam has an impact on the neutron yield. The highest yield is recorded with the target, oriented perpendicular to the beam. The neutron yield is observed to decrease with increasing β and is 3 times smaller at β=±45 degrees as compared with that at β= 0°. Such a strong angular dependence of the neutron yield could indicate a presence of narrow channels in the CVD-diamond sample, where the bulk of deuterium, trapped during the electrolysis, is concentrated. These channels could be created by the grain boundaries, which are almost vertical in the vicinity of the growth side. Another reason could be the channeling of ions in the oriented crystallites. Large neutron yield at β=0° can be explained by increased effective range of the deuterium ions inside the channels with respect to that in the diamond.